\documentstyle[amstex,prd,aps,graphicx,epsf,tighten]{revtex}

\begin{document}
\title{Cosmological Solutions in Macroscopic
Gravity}
\author{A.A. Coley\dag, N. Pelavas\dag ~and R.M. Zalaletdinov\dag\ddag}
\address{\dag\ Department of Mathematics and Statistics,\\
Dalhousie University, Halifax, Nova Scotia}
\address{\ddag\ Department of Mathematics, Statistics, and Computer Science,
\\
St. Francis Xavier University, Antigonish, Nova Scotia}

\maketitle

\begin{abstract}

In the macroscopic gravity approach to the averaging problem in cosmology, the Einstein field equations on cosmological
scales are modified by appropriate gravitational correlation terms. We present exact cosmological solutions to the
equations of macroscopic gravity for a spatially homogeneous and isotropic macroscopic space-time and find that the
correlation tensor is of the form of a spatial curvature term. We briefly discuss the physical consequences of these
results.
\end{abstract}

\noindent
[PACS: 98.80.Jk,04.50.+h]
\vskip1pc

The averaging problem in cosmology and general relativity (GR)
is of fundamental importance \cite{Ellis:1984}.
An averaging of inhomogeneous
spacetimes can lead to dynamical behaviour different
from the spatially homogeneous and isotropic Friedmann-Lemaitre-Robertson-Walker
(FLRW) model; in
particular, the expansion rate may be significantly affected \cite{Bild-Futa:1991}.
This motivated the macroscopic gravity (MG) approach to the
averaging problem in cosmology, in which
the Einstein equations on the cosmological scales with a continuous
distribution of cosmological matter
are modified by appropriate gravitational correlation correction terms \cite{Zala}.

There are a number of approaches to the averaging problem
\cite{Bild-Futa:1991,aver}. The perturbative approach
involves averaging the perturbed Einstein equations; however, a perturbation analysis
cannot provide any information about an averaged geometry.
In the space-time or space
volume averaging approach tensors, and in some cases only
scalar quantities, are averaged; this procedure is not
generally covariant hence the results are somewhat limited
and the conclusions unreliable.
In all of these approaches, in analogy with Lorentz's
approach to electrodynamics,  an averaging of the Einstein
equations is performed to obtain the averaged field equations. But
to date, with the exception of
the MG approach \cite {Zala},
no
proposal has been made about the correlation functions which should
inevitably emerge in an averaging of a non-linear theory (without which
the averaging of the Einstein
equations simply amount to definitions of the new averaged terms).

In particular, approaches to describe FLRW cosmologies as locally inhomogeneous
cosmological models utilize a 3+1
cosmological space-time splitting with non-covariant space volume averaging. The size of the
averaging space regions has been tacitly assumed to be
$\simeq $\ $100$\ Mpc, or of the order of the inverse Hubble scale.
Though many of the approaches have indicated that Friedmann's equation gets
modified by the appearance of an effective averaged energy density, there is
no definite consensus as yet on the physical status and mathematical
reliability of this important prediction on the possible dynamical law of
the Universe evolution on its largest scales.

The space-time averaging procedure adopted in MG is based on the
concept of Lie-dragging of averaging regions, which makes it valid
for any differentiable manifold with a volume $n$-form, and it has
been proven to exist on an arbitrary Riemannian space-times with
well-defined local averaged properties \cite{Zala}.  Averaging of
the structure equations for the geometry of GR brings about the
structure equations for the averaged (macroscopic) geometry and
the definitions and the properties of the correlation tensors. The
averaged Einstein equations for the macroscopic metric tensor
together with a set of algebraic and differential equations for
the correlation tensors become a coupled system of the macroscopic
field equations for the unknown macroscopic metric, correlation
tensor, and other objects of the theory. The averaged Einstein
equations can always be written in the form of the Einstein
equations for the macroscopic metric tensor when the correlation
terms are moved to the right-hand side of the averaged Einstein
equations to serve as the geometric modification to the averaged
(macroscopic) matter energy-momentum tensor. Thus, MG is a
geometric field theory with a built-in scale which is
non-perturbative and provides us with both the geometry underlying
the macroscopic gravitational phenomena and the macroscopic
(averaged) field equations  \cite{Zala}. The scale is given by the
size of the space-time averaging regions which is a free parameter
of the theory. When applied to study cosmological evolution, the
theory of MG can be regarded as a long-distance modification of
GR.

A procedure for solving the system of MG equations with one connection correlation tensor $Z^{\alpha }{}_{\beta \gamma
}{}^{\mu }{}_{\nu \sigma }$ (in brief ${\bf Z}$) is as follows {\footnote{ The MG equations are described in detail in
\cite{Zala} (wherein all terms are defined); in order to make this Letter as easy to read as possible, we shall simply
present the necessary details in a brief and compact fashion.}}. The line element for the macroscopic geometry is given
in terms of the macroscopic metric tensor $G_{\alpha \beta }$; its Levi-Civita connection coefficients and the
Riemannian curvature tensor $M^{\alpha }{}_{\beta \gamma \delta }$ can be calculated in terms of the unknown metric
functions. The components of ${\bf Z}$, perhaps with an assumption on their functional form based on symmetries and
physical conditions, can then be expressed in terms of the metric functions. The integrability conditions for the
differential equations (\ref{DZ=0}) (the ${\bf ZM}$ eqns.)
\begin{equation}
Z^{\epsilon }{}_{\beta \lbrack \gamma }{}^{\mu }{}_{\underline{\nu }\sigma
}M^{\alpha }{}_{\underline{\epsilon }\lambda \rho ]}-Z^{\alpha }{}_{\epsilon
\lbrack \gamma }{}^{\mu }{}_{\underline{\nu }\sigma }M^{\epsilon }{}_{
\underline{\beta }\lambda \rho ]}+Z^{\alpha }{}_{\beta \lbrack \gamma
}{}^{\epsilon }{}_{\underline{\nu }\sigma }M^{\mu }{}_{\underline{\epsilon }
\lambda \rho ]}-Z^{\alpha }{}_{\beta \lbrack \gamma }{}^{\mu }{}_{\underline{
\epsilon }\sigma }M^{\epsilon }{}_{\underline{\nu }\lambda \rho ]}=0,
\label{ic:DZ=0}
\end{equation}
where an underbar denotes that that index is not included in the
antisymmetrization, are solved. The system of differential
equations for ${\bf Z}$ (the ${\bf dZ}$ eqns.)
\begin{equation}
Z^{\alpha }{}_{\beta \lbrack \gamma }{}^{\mu }{}_{\underline{\nu }\sigma
\parallel \lambda ]}=0,  \label{DZ=0}
\end{equation}
where '$\parallel$' denotes covariant differentiation with respect
to the macroscopic metric, are then solved. Finally, the quadratic
algebraic conditions for $Z^{\alpha }{}_{\beta \lbrack \gamma
}{}^{\mu }{}_{\underline{\nu }\sigma ]}$ (the ${\bf ZZ}$ eqns.)
\begin{gather}
Z^{\delta }{}_{\beta \lbrack \gamma }{}^{\theta }{}_{\underline{\kappa }\pi
}Z^{\alpha }{}_{\underline{\delta }\epsilon }{}^{\mu }{}_{\underline{\nu }
\sigma ]}+Z^{\delta }{}_{\beta \lbrack \gamma }{}^{\mu }{}_{\underline{\nu }
\sigma }Z^{\theta }{}_{\underline{\kappa }\pi }{}^{\alpha }{}_{\underline{
\delta }\epsilon ]}+Z^{\alpha }{}_{\beta \lbrack \gamma }{}^{\delta }{}_{
\underline{\nu }\sigma }Z^{\mu }{}_{\underline{\delta }\epsilon }{}^{\theta
}{}_{\underline{\kappa }\pi ]}+ \\
\quad Z^{\alpha }{}_{\beta \lbrack \gamma }{}^{\mu }{}_{\underline{\delta }
\epsilon }Z^{\theta }{}_{\underline{\kappa }\pi }{}^{\delta }{}_{\underline{
\nu }\sigma ]}+Z^{\alpha }{}_{\beta \lbrack \gamma }{}^{\theta }{}_{
\underline{\delta }\epsilon }Z^{\mu }{}_{\underline{\nu }\sigma }{}^{\delta
}{}_{\underline{\kappa }\pi ]}+Z^{\alpha }{}_{\beta \lbrack \gamma
}{}^{\delta }{}_{\underline{\kappa }\pi }Z^{\theta }{}_{\underline{\delta }
\epsilon }{}^{\mu }{}_{\underline{\nu }\sigma ]}=0,  \label{ZZ=0}
\end{gather}
are solved. Upon determining the components of ${\bf Z}$, the
gravitational stress-energy tensor $T_{\beta }^{\alpha {\rm
(grav)}}$ of MG, defined by
\begin{equation}
(Z^{\alpha }{}_{\mu \nu \beta }-\frac{1}{2}\delta _{\beta }^{\alpha }Q_{\mu
\nu })\bar{g}^{\mu \nu }=-\kappa T_{\beta }^{\alpha {\rm (grav)}},
\label{RicZ=T(grav)}
\end{equation}
is determined (where $Z^{\alpha }{}_{\mu \nu \beta } \equiv 2Z^{\alpha }{}_{\mu \epsilon }{}^{\epsilon }{}_{\nu \beta
}$, $Z^{\epsilon }{}_{\mu \nu \epsilon } \equiv Q_{\mu \nu }$). The averaged Einstein equations {\footnote{The
macroscopic field equations (\ref{M=<t>+Z}) are written in the form of the Einstein equations of GR, with a 'modified'
stress-energy tensor consisting of the averaged microscopic stress-energy tensor $\langle {\bf t}_{\beta }^{\alpha {\rm
(micro)}}\rangle$ and an additional effective stress-energy tensor $T_{\beta }^{\alpha {\rm (grav)}}$
(\ref{RicZ=T(grav)}) arising from the correlation tensor ${\bf Z}$ \cite{Zala}.}}
\begin{equation}
G^{\alpha \epsilon }M_{\epsilon \beta }-\frac{1}{2}\delta _{\beta }^{\alpha
}G^{\mu \nu }M_{\mu \nu }=-\kappa \langle {\bf t}_{\beta }^{\alpha {\rm
(micro)}}\rangle
-\kappa T_{\beta }^{\alpha {\rm (grav)}} \label{M=<t>+Z}
\end{equation}
are then solved
for the unknown metric functions, assuming
(for example)
that the averaged microscopic stress-energy tensor $\langle
{\bf t}_{\beta }^{\alpha {\rm (micro)}}\rangle $ is of a perfect fluid
form.

Given a macroscopic metric $G_{\alpha\beta}$, the calculational procedure is to seek a solution ${\bf Z}$ satisfying
the ${\bf ZM}$ , ${\bf dZ}$ and ${\bf ZZ}$ eqns.  By making extensive use of GRTensorII \cite{grtensor} and Maple, the
first step is to define the connection correlation tensor with its rank and symmetries.  In practice, a file is created
for a rank 6 tensor ${\bf Z}$ possessing no symmetries, the symmetries on ${\bf Z}$ are then imposed by solving systems
of algebraic equations.  The choice of metric at this stage is irrelevant.  Although solving the ${\bf ZZ}$ eqns. does
not involve the metric, we have found it convenient to solve this equation last; since it is quadratic, many solution
sets will arise and only after ${\bf Z}$ has been constrained either by ${\bf ZM}$ and ${\bf dZ}$ , or any other
additional assumptions on ${\bf Z}$, is there a possibility of solving ${\bf ZZ}$ computationally.  To each solution
set of ${\bf ZZ}$ there will be a corresponding $T_{\beta}^{\alpha \mathrm{(grav)}}$. A typical worksheet begins with
the loading of a macroscopic metric and the connection correlation tensor. It is useful to have a set of the
independent components of ${\bf Z}$, this is easily done by looping through all components of ${\bf Z}$. We begin by
defining a rank 8 tensor corresponding to the ${\bf ZM}$  eqns. These algebraic equations are then solved for the
independent components of ${\bf Z}$ and the solutions are substituted back into ${\bf Z}$.  It is easily checked that
${\bf Z}$ now satisfies the ${\bf ZM}$  eqns.  Next we define a rank 7 tensor corresponding to the ${\bf dZ}$  eqns. If
a solution of these differential equations for the independent components of ${\bf Z}$ can be found, it can then be
substituted back into the ${\bf Z}$ tensor. In most of the cases considered, we have found no great computational
difficulty in solving the ${\bf ZM}$ and ${\bf dZ}$  eqns. using Maple. At this point the number of independent
components of ${\bf Z}$ left unspecified by the ${\bf dZ}$ eqns. can be computed. To define the ${\bf ZZ}$  eqns. we
define six rank 6 tensors, each corresponding to a term of the ${\bf ZZ}$  eqns. fully contracted with the Levi-Civita
tensor over the anti-symmetrized indices. These tensors are calculated individually then summed to give a rank 6 tensor
corresponding to the ${\bf ZZ}$  eqns. As above, this tensor can be calculated and its components stored in a set. We
then solve for the remaining independent components of ${\bf Z}$. Since multiple solutions will be obtained it is
necessary to define and calculate multiple copies of the ${\bf Z}$ tensor. There are many variations to the outline
given above, depending on the form of the metric and the assumptions on the components of ${\bf Z}$. For example,
assuming that the components of ${\bf Z}$ are all constants in an appropriate form, the ${\bf ZM}$  and ${\bf dZ}$
eqns. amount to algebraic equations, thus eliminating the need to solve any differential equations.

Let us consider a flat spatially homogeneous, isotropic macroscopic
FLRW space-time with conformal time $
\eta $
\begin{equation}
ds^{2}=a^{2}(\eta )(-d\eta ^{2}+dx^{2}+dy^{2}+dz^{2}),  \label{rw-flat}
\end{equation}
where $d\eta =a^{-1}(t)dt$ with a cosmological (coordinate) time
$t$ and $ a^{2}(t)$ is an unknown function of the scale factor. It
is necessary to make an ansatz for the functional form of the
components of ${\bf Z}$ on the basis of symmetries and physical
conditions of the macroscopic geometry. The most natural condition
on ${\bf Z}$ compatible with the structure of macroscopic
space-time (\ref{rw-flat}) is to require all of its components be
constant
\begin{equation}
Z^{\alpha }{}_{\beta \gamma }{}^{\mu }{}_{\nu \sigma }={\rm
const}. \label{Z6=const}
\end{equation}
%%(i.e., we assume that the MG correlations do not change in time and space).

Upon solving the ${\bf ZM}$ and ${\bf dZ}$ eqns., using a Maple built-in algebraic system solver and requiring
real-valued solutions, we are left with a number of independent components in ${\bf Z}$. Solving the ${\bf ZZ}$ eqns.
then yields  a number of solutions (with a small number of non-vanishing real-valued components of ${\bf Z}$), each  of
which gives $T_{x}^{x \mathrm{(grav)}} =T_{y}^{y \mathrm{(grav)}}=T_{z}^{z \mathrm{(grav)}}=\frac{1}{3}T_{t}^{t
\mathrm{(grav)}}$, where $T_{x}^{x \mathrm{(grav)}}=-\beta/a^2(t)$ and $\beta$ is a linear combination of the non-zero
constant components of ${\bf Z}$ (different combinations corresponding to different solutions; e.g., $\beta = -12
Z^{3}{}_{23}{}^{3}{}_{32}$ in three particular exact solutions with a single independent component of ${\bf Z}$). In
all cases the MG stress-energy tensor has the form of a perfect fluid with $\rho_c=\beta /\kappa a^{2}(t)$ and
$p_c=-\rho_c/3$ (i.e., $\gamma = 2/3$). After transforming from conformal time $\eta $ to cosmological time $t$, we
obtain the averaged Einstein equations
\begin{equation}
\left( \frac{\dot{a}}{a}\right) ^{2}=\frac{\kappa \rho }{3}+\frac{\kappa \beta }{
3a^{2}(t)},  ~~
2\frac{\ddot{a}}{a}+\left( \frac{\dot{a}}{a}\right) ^{2}=-\kappa p+\frac{
\kappa \beta }{3a^{2}}.  \label{macro-law2}
\end{equation}

Thus, the averaged Einstein equations for a flat spatially
homogeneous, isotropic macroscopic space-time geometry has the
form of the Einstein equations of GR for a open {\footnote{In
principle, without imposing any further conditions, the curvature
can be positive or negative. However, if the energy density
$\rho_c$ of the MG field is positive, then
 $trace(T_{\beta }^{\alpha {\rm (grav)}}{})=-2\beta /a^{2}(t) <0$
(i.e., ${\bf T}^{\mathrm{(grav)}}$ is a {\em negative} curvature term), which means from the physical point of view
that the macroscopic gravitational energy is the binding energy of the Universe.}} spatially homogeneous, isotropic
space-time geometry (where the correlation tensor is of the form of a spatial curvature term, with $k = -
\beta/3\kappa$). In all cases (i.e., calculations in which different assumptions on the form of ${\bf Z}$ are made),
solutions always give rise to a spatial curvature term.  Indeed, assuming only spatial correlations (i.e., assuming
that all components of ${\bf Z}$ with at least one t index must vanish), it can be shown that ${\bf
T}^{\mathrm{(grav)}}$ must be of the form of spatial curvature. This is the main result of this Letter; namely,  {\em
for a flat FLRW geometry the MG correlations are of the form of a spatial curvature tensor term}. In further
experimentation, in some non-flat spatially homogeneous, isotropic macroscopic models we also found evidence that ${\bf
T}^{\mathrm{(grav)}}$ is of the form of a curvature term.

There are a number of important physical consequences of these results.
In MG, in a flat spatially homogeneous and
isotropic macroscopic
space-time, the
correlation tensor and the averaged cosmological matter distribution taken
as a perfect fluid has the cosmological dynamical equations (\ref{macro-law2}).
This implies that the macroscopic (averaged) cosmological
evolution in a flat Universe is governed by the dynamical evolution
equations for an open Universe, which makes it necessary to
reconsider the standard cosmological interpretation and
the
treatment of the observational data. If the underlying macroscopic
space-time has positive spatial curvature (as suggested by
recent observations), then we would obtain a cosmological model
which is closed on local scales, but as a result of the
MG correlations behaves dynamically on
macroscopically large scales as a flat model, which might have considerable
physical implications. Finally, if positive spatial curvature correlations
are permitted, then cosmological models which act like
an Einstein static model on the largest scales are possible even for
models with zero or negative curvature on small scales. Thus we have the interesting,
but highly conjectural, possibility that since at late times (and on the largest scale)
${\bf T}^{\mathrm{(grav)}}$
(a curvature term) will dominate the dynamics,
the correlations might stabilize
the  Einstein static model. This may be of potential importance
since current observations perhaps indicate that the universe is
marginally closed and due to the current interest
in the emergent universe scenario
in which the universe is
positively curved and initially in a past eternal Einstein static
state that eventually evolves into a subsequent inflationary phase
\cite{Ellis-Maartens}.

Let us discuss the potential significance of these results in a little more detail.
Observations are usually interpreted as showing that
the Universe is flat, currently accelerating
and indicating the existence of dark matter and dark energy \cite{Weinberg}.
As noted  earlier, inhomogeneities can affect the dynamics
and may significantly affect the expansion rate \cite{Bild-Futa:1991}. It has been
suggested that back-reaction from
inhomogeneities smaller than the Hubble scale could
explain the apparently observed accelerated expansion
of the universe today or negate the need for dark energy
in a realistic inhomogeneous  universe.
Indeed, it has been argued that the cosmological constant can
be reduced to a very small value by
back-reaction effects in an expanding space-time \cite{Nobbenhuis}.
For example, gravitational waves propagating in a background spacetime
will affect the dynamics of this background.
The back-reaction for scalar gravitational perturbations, which can be
described by an effective energy-momentum tensor, was studied in \cite{Brandenberger}.
It was found  that the equation of state of the dominant infrared
contribution to the energy-momentum tensor which
describes back-reaction can take the form of a negative cosmological constant.
This has led to the speculation that gravitational back-reaction may
lead to a dynamical mechanism for the cancellation of a bare cosmological constant.
However, it is not clear whether this approach is consistent
and whether the effects are indeed physical. For example, averaging over
a fixed time slice, the spatially averaged value of the expansion will not
be the same as the expansion rate at the averaged value of time,
because of the non-linear nature of the expansion.

What is needed is a correct averaging procedure that does not depend on any assumptions
regarding the nature of the perturbations.
The MG method described here is an exact approach;
no approximations have been made (i.e., no higher order terms have been dropped).
In this approach inhomogeneities affect the dynamics on large scales
through correction terms (and, in this sense, is different to back-reaction effects
which are pure non-linear effects of the gravitational field via perturbations).
Moreover, averaging entails a scale dependence, which depends on the spatial scale
over which averages are taken. This averaging scale
is assumed to be of the order of the inverse Hubble scale, and thus any terms
(e.g., a cosmological constant or a curvature term)
appearing in the correlation tensor must be related to the inverse Hubble scale.
For example, the natural length scale
of any cosmological constant introduced by averaging would be of the order of the inverse
Hubble scale squared. This would therefore give a natural possible resolution of the
coincidence problem \cite{Nobbenhuis}.
Unfortunately, to date we have not been able to solve the MG equations
to find a solution with correction terms that
may account for the present day acceleration.
However, a spatial curvature
correction arises naturally and, as noted earlier,
correction terms change the
interpretation of observations so that they need to be accounted for
carefully to determine if they may be consistent
with a decelerating universe.

In addition, superhorizon fluctuations (whose origin is in inflation)
affect classical dynamics as measured by local observers (since perturbations
affect the expansion rate in a universe with a flat FLRW background).
Recently it has been proposed that superhorizon
perturbations could explain the present-day accelerated  acceleration \cite{Kolb:2005}.
However, in \cite{Geshnizjani:2005} it was claimed that the effect
proposed in \cite{Kolb:2005} amounts to
a simple renormalisation of the spatial curvature
(essentially a new scale factor can be defined so that the metric
looks like a FLRW metric with a curvature term), and thus cannot account for
negative deceleration; indeed, a proper
accounting of all perturbative terms as well as
more general arguments suggest that the superhorizon modes
do not lead to acceleration \cite{Flanagan:2005}.
In  further work \cite{Rasanen:2005} the relation between backreaction (and especially
the effective scale factor presented in
\cite{Kolb:2005}) and spatial
curvature using exact equations which do not rely
on perturbation theory was studied in more detail; it was found
that although the effect does not simply reduce to spatial curvature,
acceleration results but is accompanied by growth of spatial curvature
to an extent that is likely to be incompatible with the CMB data.

{\em Acknowledgements}. We would like to thank Robert van den Hoogen
for helpful discussions and calculations. This work was supported, in
part, by NSERC. RZ acknowledges the support by a
research grant from St. Francis Xavier University attached to his W.F. James
Chair.

\end{document}